\documentclass[12pt]{article}  
\usepackage{a4,latexsym}
\usepackage{setspace}
\usepackage{abstract}
\usepackage[T1]{fontenc}
\usepackage[colorlinks,allcolors=blue]{hyperref}
\usepackage{tabularx}
\usepackage{graphicx}
\usepackage{enumerate}
\usepackage{amssymb}
\usepackage{amsmath}
\usepackage{caption}
\usepackage{natbib}
\usepackage{booktabs}
\usepackage{float}
\usepackage[top=1.5in, bottom=1.1in, left=1in, right=0.875in]{geometry}
\usepackage{amsfonts,array}
\usepackage{epsfig, natbib, multirow}
\usepackage{mathtools, subfig, ragged2e, colortbl, enumitem, soul}
\usepackage{ulem}

\usepackage{xcolor}

\usepackage{color}

\numberwithin{equation}{section}

\begin{document}
\normalem

\title{
	\vspace{-45mm}
	\fbox{
		\begin{minipage}{\textwidth}%
			\begin{singlespace}
				\small{  
					\vspace{-0.05mm}
					This is the peer reviewed version of the following article: Ugba, E. R. and Gertheiss, J. (2023). A Modification of McFadden's $R^2$ for Binary and Ordinal Response Models.{ \it Communications for Statistical Applications and Methods},{ \it 30}, 49--63, which has been published in final form at
					\href{https://doi.org/10.29220/CSAM.2023.30.1.049}{https://doi.org/10.29220/CSAM.2023.30.1.049}.
					This manuscript version is made available under the arXiv's
					\href{https://arxiv.org/licenses/nonexclusive-distrib/1.0/license.html}{Non-exclusive license to distribute}.}
			\end{singlespace}
		\end{minipage}
	} \vspace{5mm}\\
	\textbf{A Modification of McFadden's $R^2$ for Binary and Ordinal Response Models
	}
} 
   
\maketitle

\begin{center}
\textbf{Abstract}
\end{center}
A lot of studies on the summary measures of predictive strength of categorical response models consider the likelihood ratio index (LRI), also known as the McFadden-$R^2$\!, a better option than many other measures. We propose a simple modification of the LRI that adjusts for the effect of the number of response categories on the measure and that also rescales its values, mimicking an underlying latent measure. The modified measure is applicable to both binary and ordinal response models fitted by maximum likelihood. Results from simulation studies and a real data example on the olfactory perception of boar taint show that the proposed measure outperforms most of the widely used goodness-of-fit measures for binary and ordinal models. The proposed $R^2$ interestingly proves quite invariant to an increasing number of response categories of an ordinal model.

\vspace{1cm}

\newpage

\section{Introduction}
Determining the predictive strength of a categorical response model is neither an easy nor straightforward task as it is the case with linear models. The coefficient of determination $(R_{(\text{ols})}^{2})$ in ordinary least squares regression provides an intuitive and very well specified measure of fit, but is largely inapplicable to discrete response models. Although several $R^2$-like measures, popularly known as `Pseudo-$R^{2}$'\!, have been suggested in the literature for the assessment of the predictive strength of categorical models (see for example, \cite{mcFadden_conditional_1974, cox_analysis_1989, nagelkerke_note_1991, mcKelvey_statistical_1976, heinzl_pseudo_2003, tjur_coefficients_2009, zhang_coefficient_2017, piepho_coefficient_2019}), no meaningful consensus has yet been reached on which of those performs best in empirical studies. A couple of studies in the past seem to favor McFadden's $R_\text{(mf)}^{2}$, considering its easy computation, intuitive interpretation, base-rate stability in binary models and possible information theory interpretation (see, \cite{hauser_testing_1978, windmeijer_goodness_1995, menard_coefficients_2000}). Nevertheless, despite all its appealing features, $R_\text{(mf)}^{2}$ still has some drawbacks that render its use questionable, particularly in ordinal response models. As \cite{long_regression_1997} observed, there is no clear interpretation of values other than zero and one, in other words, values between these two ranges seem somewhat arbitrary since there is no meaningful way to determine if they are large or small. Moreover, \cite{hagle_goodness_1992} observed that $R_\text{(mf)}^{2}$ significantly underestimates $R_\text{(ols)}^{2}$ of an underlying continuous model.

We propose a simple modification of $R_\text{(mf)}^{2}$ that addresses its key limitations, making it useful for both binary and ordinal models obtained via the maximum likelihood. The proposed measure is presented in Section~\ref{sec:Modified} after a short discussion on the latent variable motivation of binary/ordinal models in Section~\ref{Sec:Latent}. A simulation study and an empirical application are  presented in Sections~\ref{Sec:Sim} and ~\ref{Sec:Empirical}, respectively. Section~\ref{Sec:Disc} concludes.

\section{Latent Variable Motivation}\label{Sec:Latent}
Given an ordinal response $y_i$ for subject $i$ = $1,\ldots,n$, with potential values $1,\ldots,r$, consider a continuous underlying latent variable $\tilde{y}_{i}$ with the following generating function:
\begin{align}
	\tilde{y}_{i} = \boldsymbol{x}_i^\top\tilde{\boldsymbol{\beta}} + \epsilon_i, \quad i=1,\ldots,n,
\end{align}
where $\boldsymbol{x}_i$ is a vector of covariates, $\tilde{\boldsymbol{\beta}}$ a vector of regression parameters and $\epsilon_i$ an error term. Suppose $-\infty$ = $\tau_0<\tau_1<\cdots<\tau_r$ = $\infty$ are cut-points on $\tilde{y}_{i}$ such that the observed response $y_{i}$ satisfies the threshold model,
\begin{align*}
	y_i=j \Leftrightarrow \tau_{j-1} < \tilde{y}_{i} <\tau_{j},
\end{align*}
with $j$ = $1,2,\ldots,r$. For the error term $\epsilon_{i}$, typically a normal or logistic distribution is assumed, leading to a so-called cumulative probit or logit model, respectively; see, e.g., \cite{agresti_categorical_2002} for details. Models obtained through this means are said to be latent variable motivated, and could possibly reference the originating model. As a consequence, a common criterion when assessing the goodness-of-fit of such a model is that the Pseudo-$R^{2}$ used to measure its predictive strength should be as close as possible to the $R_\text{(ols)}^{2}$ of the underlying continuous model; see, for example, \cite{hagle_goodness_1992}; \cite{windmeijer_goodness_1995}; \cite{veall_pseudo_1992}. The $R_\text{(ols)}^{2}$ is the popular coefficient of determination in the classical linear model calculated through
\begin{align}\label{r2_ols}
	R_\text{(ols)}^{2}=1- \frac{\sum_i{\left(\tilde{y}_i - \hat{\tilde{y}}_i\right)^{2}}}{\sum_i{\left(\tilde{y}_i - \bar{\tilde{y}}\right)}^{2}},
\end{align}
where $\hat{\tilde{y}}_i$ denotes the fitted (latent) response of subject $i$ using the estimated parameters $\hat{\tilde{\boldsymbol{\beta}}}$ and the explanatory variables $\boldsymbol{x}_i$, and $\bar{\tilde{y}}$ denotes the sample (arithmetic) mean of the $\tilde{y}_i$. This measure has some interesting properties that makes it widely useful. As noted in \cite{rao_linear_1973},
\begin{enumerate}[leftmargin=\parindent,align=left,labelwidth=\parindent,labelsep=0pt]
	\item[1.] It has an easy and intuitive interpretation as the proportion of the variance explained by the model.
	\item[2.] It lies between 0 and 1.
	\item[3.] It is dimensionless, i.e., it is independent of all units of measurement from the variables, and
	\item[4.] it is independent of sample size.
\end{enumerate}

The literature is inundated with several analogs to $R_\text{(ols)}^{2}$ designed to achieve some of the above listed criteria for generalized linear models and their extensions. One of such is the so-called likelihood ratio index (LRI), also known as the McFadden's $R_\text{(mf)}^{2}$ \citep{mcFadden_conditional_1974, maddala_limited_1983}, and may be expressed as follows;
\begin{align}
	R_\text{(mf)}^{2} = 1 - \frac{\ell_p}{\ell_0},
\end{align}
where $\ell_p$ is the (maximum) log-likelihood of the full model (with $p$-predictors) and $\ell_0$ is the log-likelihood of the (null) model with intercept alone. With some algebra one has $R_\text{(mf)}^{2}$ = $G/{(-2\ell_0)}$, where $G$ = $-2[\ell_0 - \ell_p]$ is the model chi-square statistics and $-2\ell_0$ the minus two log-likelihood statistic of the null model.

According to \cite{hosmer_applied_1989}, the latter quantity is identically equal to the sum of squared errors in the ordinary least squares (OLS) null model, i.e., it captures the error variation of the model with only the intercept present \citep{nagelkerke_note_1991, menard_coefficients_2000}. Thus, similar to the $R_\text{(ols)}^{2}$ which is interpreted as the proportional reduction in the error sum of squares, the $R_\text{(mf)}^{2}$ is more or less considered the proportional reduction in the $-2$ log-likelihood statistic \citep{menard_coefficients_2000}. This makes it a bit more intuitive than most competing measures and also widely reported in empirical studies. However, in addition to
the limitations of $R_\text{(mf)}^{2}$ already mentioned in the Introduction, one crucial but often ignored question is how $R_\text{(mf)}^{2}$ changes with increasing number of response categories of an ordinal model. In other words, how does a regrouped response category (merging or splitting) given the same dataset affect $R_\text{(mf)}^{2}$? Going by the invariance property of cumulative models motivated through an underlying, latent variable, the same parameters occur for the effects regardless of how the cut-points discretize the continuous scale \citep{agresti_categorical_2002}. Consequently, an adequate summary measure of predictive strength of categorical models should yield consistent conclusion with the same dataset irrespective of changes in the number of response categories. This, however, is not the case with $R_\text{(mf)}^{2}$ given in ~(\ref{r2_ols}). In general, the quantity
\begin{align}\label{gamma_r}
	\gamma_{r} = \frac{\ell_{p}\left(r\right)}{\ell_{0}\left(r\right)}, \quad r = 2,3,\dots,
\end{align}
capturing the amount of likelihood explained in a given model happens to depend on $r$. More specifically, if sample size $n$ remains fixed and $r$ increases, $\gamma_{r}$ approaches 1 (it's upper limit) because $\ell_0(r)$ tends towards the value of $\ell_{p}(r)$ due to the increasing number of threshold parameters/cut-points. As a consequence, $R_\text{(mf)}^{2}$ decreases with increasing $r$. One implication of this is that two researchers with different choice of response categories for studying the same predictor effect may eventually end up with two different conclusions. Hence, there is apparently some need for a correcting factor on the likelihood ratio index in case of ordinal response models.

\section{The Modified Measure}\label{sec:Modified}

\begin{figure}[t!]
	\centering
	\includegraphics[width=100mm]{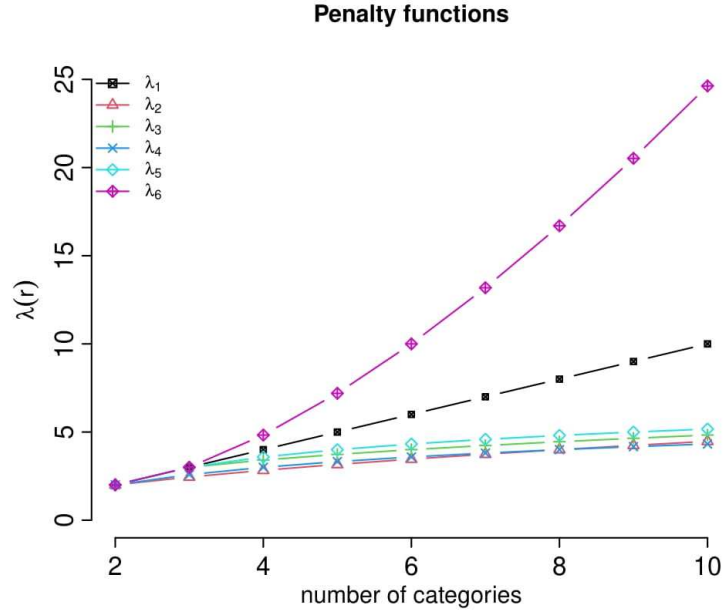}
	\vspace{1ex}
	\caption{\label{fig:penalty} Candidate penalty functions for the modified McFadden measure plotted against an increasing number of response categories $(r = 2,3,\dots,\!10)$.}
	\vspace{1.5ex}
\end{figure}

As already seen, the dependency of $R_\text{(mf)}^{2}$ on the number of response categories $r$, especially in ordinal models motivated by an underlying continuous model, is hardly desirable for a supposedly good measure of fit. \cite{Agresti_applying_1986} made a similar observation about so-called entropy-based measures which include $R_\text{(mf)}^{2}$, suggesting the need for an appropriate correction that could account for the effect of $r$ on such measures. Moreover, \citep{Veal_pseudo_1996} argue that in terms of ordinal models, the most useful R-squared is one that is most comparable to the $R^2$ on the underlying latent variable model, i.e., $R_\text{(ols)}^{2}$. Thus, to redress the effect of $r$ on the likelihood ratio index, we propose the use of a stabilizing exponential penalty on $\gamma_{r}$. Assuming the following penalized likelihood ratio index \citep{ugba_augmented_2018}:
\begin{align}\label{R2r}
	R_{\left(r\right)}^{2} = 1- \gamma_{r}^{\lambda\left(r\right)}, \quad r=2,3, \ldots,
\end{align}
\begin{figure}[t!]
	\centering
	\includegraphics[scale=0.3]{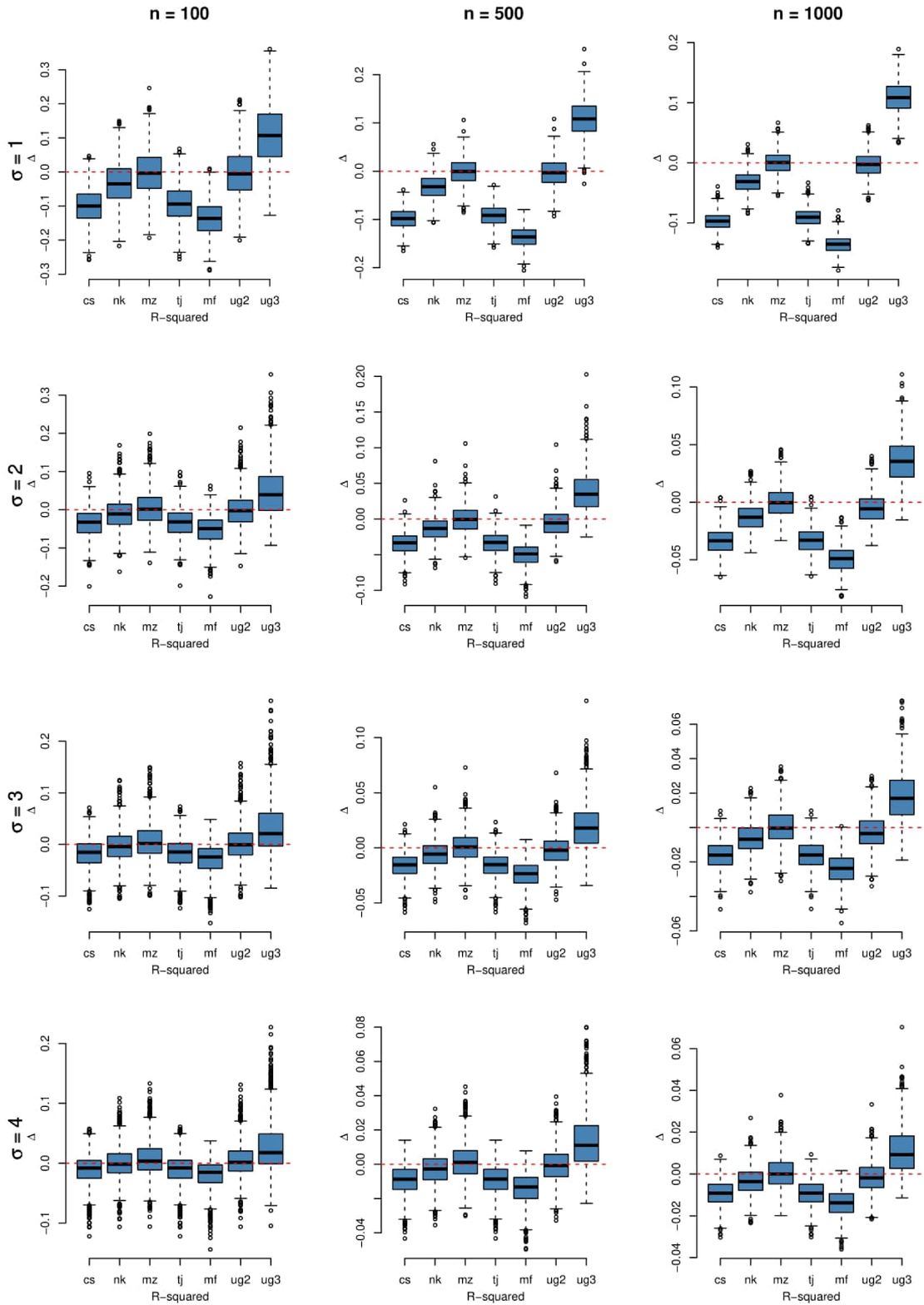}
	\vspace{3mm}
	\caption{\label{error1}The error (i.e., deviations from $R_\text{(ols)}^{2}$) of R-squared measures obtained from a thousand replications of the binary probit model, with the three columns denoting the three sample sizes considered and the rows the four different standard deviation specifications. Abbreviated measures include: Cox \& Snell (cs), Nagelkerke (nk), McKelvey \& Zavoina (mz), Tjur (tj), McFadden (mf) and two different specifications of the modified measure~(ug2 and ug3).}
	\vspace{1.5ex}
\end{figure}

where $\lambda(r)$ is a strictly positive penalty function that is monotonically increasing in $r$.
Here, we will consider and evaluate penalties with different functional shapes, including a direct one-to-one mapping of increasing number of categories, and some non-linear (mostly) concave and convex shaped penalties. Specifically, we consider the following six candidate penalty functions:
\begin{align}\label{penalty}
	\lambda_{1} \left(r\right) &= r,\nonumber\\
	\lambda_{2} \left(r\right) &= \sqrt{2r}, \nonumber\\
	\lambda_{3} \left(r\right) &= 2 + \sqrt{r-2}\ , \\
	\lambda_{4} \left(r\right) &= \log_{2}{2r} = 1 + \log_2{r}, \nonumber\\
	\lambda_{5} \left(r\right) &= 2 + \log_{2}\left(r-1\right), \nonumber\\
	\lambda_{6} \left(r\right) &= 2 + \left(r-2\right)^{\frac{3}{2}}\!. \nonumber
\end{align}

\begin{figure}[t!]
	\centering
	\includegraphics[scale=0.3]{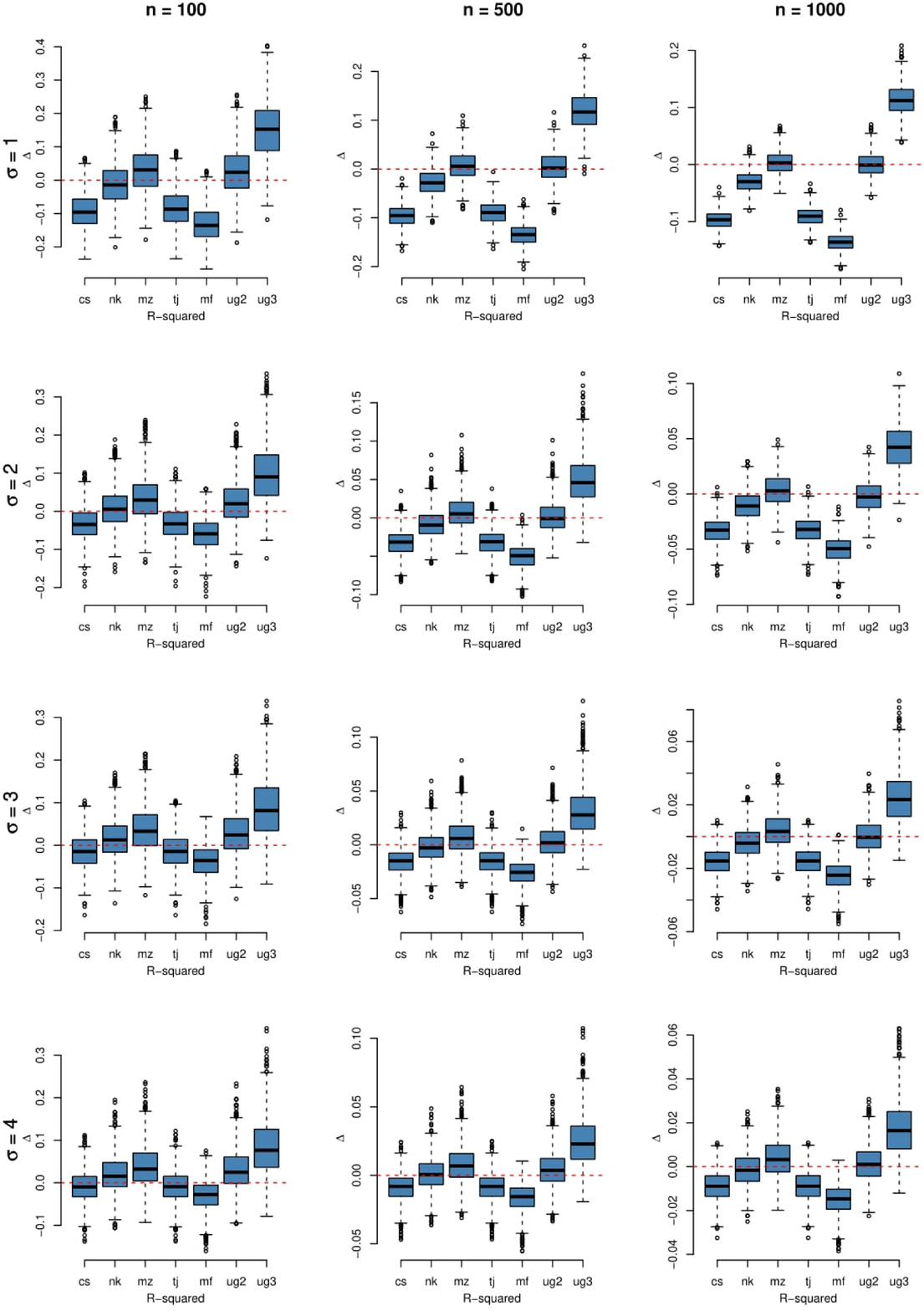}
	\vspace{3mm}
	\caption{\label{error2}The error (i.e., deviations from $R_\text{(ols)}^{2}$) of R-squared measures obtained from a thousand replications of the binary probit model with five additional, purely noise predictors. The three columns denote the three sample sizes considered and the rows the four different standard deviation specifications. Abbreviated measures include: Cox \& Snell (cs), Nagelkerke (nk), McKelvey \& Zavoina (mz), Tjur (tj), McFadden (mf) and two different specifications of the modified measure~(ug2 and ug3).}
	\vspace{1.5ex}
\end{figure}
These functions are shown in \mbox{Figure~\ref{fig:penalty}} against an increasing number of response categories $r$. The various formulations represent penalties with both high and low impact on the likelihood ratio index. The first penalty $\lambda_{1} (r)$ provides an identity mapping, $\lambda_{2} (r)$ to $\lambda_{5} (r)$ are concave shaped penalty functions, while $\lambda_{6} (r)$ provides a convex shaped penalty. Here, all functions start at $\lambda(2)$ = $2$ for binary regression, which worked well in preliminary studies (Ugba and Gertheiss, 2018), but will also be further investigated/illustrated in Section~\ref{Sec:Sim} below. Although not included \mbox{in~(\ref{penalty})}, a special (yet extreme) case of \mbox{(\ref{R2r})} results in having constant $\lambda_{0} (r)$ = $1$ for all $r$, which collapses the penalized measure to the original non-penalized likelihood ratio index (i.e., $R_\text{(mf)}^{2}$). This implies that both measures would share some properties. Indeed, with $|\ell_{p}(r)| \leq |\ell_{0}(r)|$ in fully identified likelihood estimated models, the ratio $\gamma_{r}$ given in~(\ref{gamma_r}) always lies in the interval [0, 1], meaning that for the modified $R_{(r)}^{2}$ from~(\ref{R2r}) $0 \le R_{(r)}^{2} \le 1$ holds as well. Moreover, since typically $0 < \gamma_{r} < 1$, the fact that each of the considered $\lambda(r)$ candidates is strictly monotone means that it will countervail the effect that $R_\text{(mf)}^{2}$ tends to decrease for increasing $r$. Finally, $R_{(r)}^{2}$ ideally approximates the underlying $R_\text{(ols)}^{2}$. In summary, any of the  penalties in (\ref{penalty}) resulting in a stable $R^2$-value across a varying number of ordinal response categories, $r$, and also approximating $R_\text{(ols)}^{2}$ closely, could be considered a useful penalty function. In what follows, those two aspects will be evaluated for the candidate functions in numerical experiments.

\section{Simulation Study}\label{Sec:Sim}
We present an analysis of simulated binary and ordinal response data given a variety of data generation specifications. First, the continuous underlying latent variable, $\tilde{y}_{i}$, was obtained under two different covariate settings as follows:
\begin{enumerate}[label=(\alph*),leftmargin=\parindent,align=left,labelwidth=\parindent,labelsep=0pt]
	\item \textbf{Single~distribution}, where $\boldsymbol{x}_i$ = $(x_{i1}, x_{i2})^\top$, $i$ = $1,\ldots,n$, with both variables taken from i.i.d $U(0,1)$, $n$ the number of observations and the fixed regression parameters $\boldsymbol{\tilde{\beta}}$ = $(\beta_{0}, \beta_{1}, \beta_{2})^\top$ equal to 0, 1, and 2, respectively.
	
	\item \textbf{Mixed} \textbf{ distribution}, where $\boldsymbol{x}_i$ = $(x_{i1},\ x_{i2},\ x_{i3},\ x_{i4},\ x_{i5})^\top$\!, $i$ = $1,\ldots,n$, with the first three variables taken from i.i.d $N(0,1)$ and the remaining two from i.i.d $U(0,1)$. The fixed regression parameters $\boldsymbol{\tilde{\beta}} = (\beta_{0},\ \beta_{1},\ \beta_{2},\ \beta_{3},\ \beta_{4},\ \beta_{5})^\top$ are equal to 0, $-$1/3, $-$2/3, $-$1, 1, and 2, respectively.
\end{enumerate}

The disturbance term, $\epsilon_{i}$ was taken from $N(0,\sigma)$ in both settings, with $\sigma$ denoting the standard deviation. The entire experiment was conducted with three different sample sizes $n$ = $100, 500, 1,\!000$ and four different specifications of the standard deviation $\sigma$ = 1,2,3,4. \!A thousand replications of datasets $\{(\tilde{y}_{i}, \boldsymbol{x}_i),$ $i=1,\ldots,n\}$ were obtained and response values were subsequently discretized $(r \geq 2)$ with equidistant cut-points (equal quantiles), resulting in a thousand replications of $\{(y_{i}, \boldsymbol{x}_i),$ $i=1,\ldots,n\}$. The underlying $R_\text{(ols)}^{2}$ was obtained from the continuous latent models (fitted with the \texttt{lm()} function from the \texttt{stats} R-package \citep{R_language_2022}). We move on to present the results from the first simulation setting (single predictor distribution, binary response).
\begin{figure}[t!]
	\centering
	\includegraphics[scale=0.31]{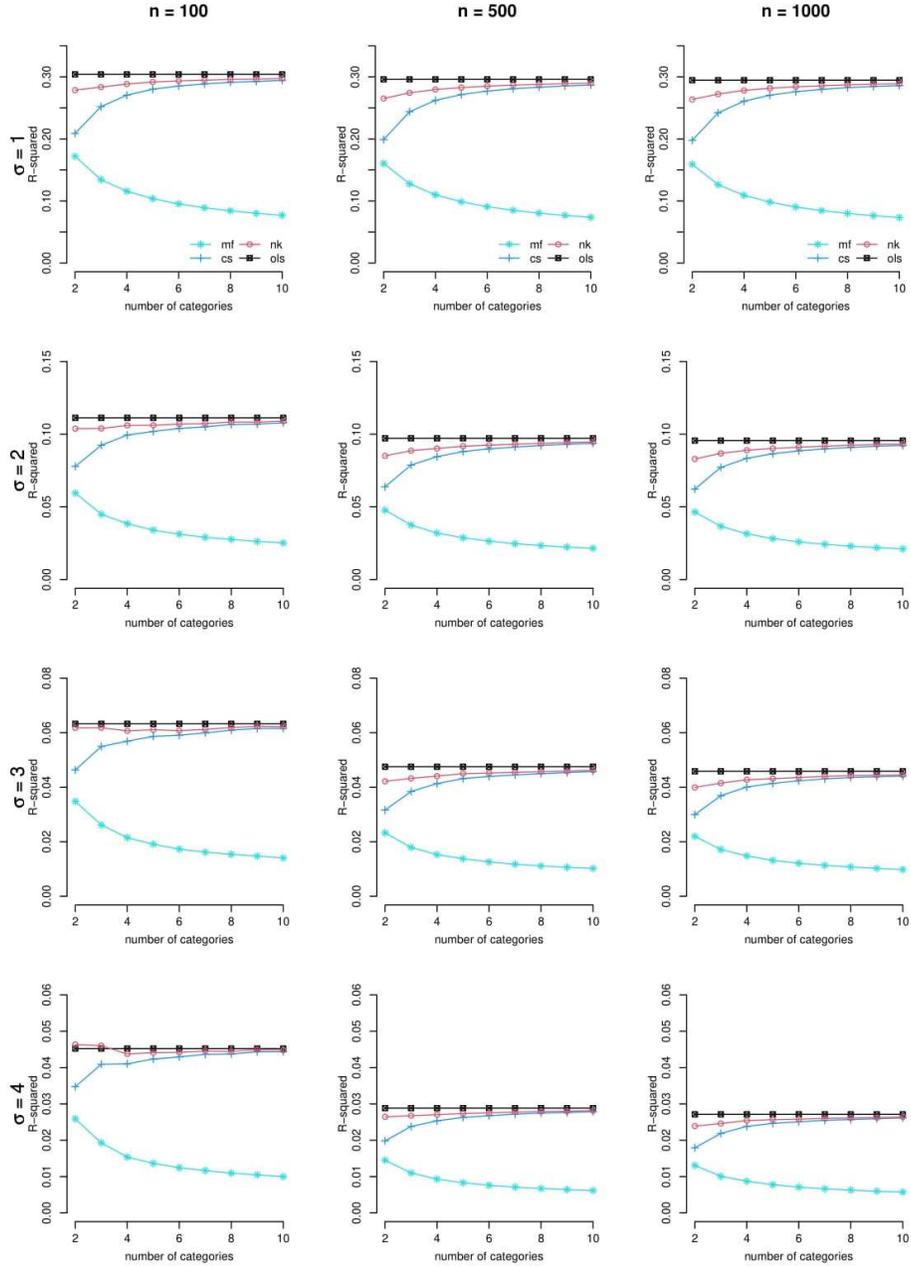}
	\vspace{3mm}
	\caption{\label{ordinal1}\!R-squared of multi-categorical response models against increasing number of response category $(r = 2, 3, \dots, 10)$, comparing existing measures with the underlying measure (ols). Rows and columns respectively denote the four different specifications of standard deviation of $\tilde{y}_{i}$  and the three sample sizes. Abbreviated measures include: McFadden (mf), Cox \& Snell (cs), Nagelkerke (nk), and the underlying measure (ols).}
	\vspace{1.5ex}
\end{figure}

\subsection{Results from binary models}\label{Sec:Binary}
The modified $R_{(r)}^{2}$ was obtained from binary probit models built with the simulated categorical/binary response, $y_{i}$, and the predictors used in the underlying model (fitted with the \texttt{glm()} function from the \texttt{stats} R-package). In the binary case, i.e., $r$ = 2, each $\lambda(r)$ candidate considered in \mbox{(\ref{penalty})} evaluates to 2. So to create a further rival penalty, we consider an additional instance where the penalty equals something else, say $\tilde{\lambda}(2)$ = 3. Moreover, to compare the modified measure with existing measures, McFadden's $R_\text{(mf)}^{2}$ together with other commonly used R-squared measures for categorical models were also obtained. These include the Cox \& Snell's $R_\text{(cs)}^{2}$ and its corrected version, popularly known as the Nagelkerke's $R_\text{(nk)}^{2}$ (see, Cox and Snell, \cite{cox_analysis_1989}; \cite{nagelkerke_note_1991}). Also obtained were the McKelvey $\&$ Zavoina $R_\text{(mz)}^{2}$ \citep{mcKelvey_statistical_1976}, and the Tjur $R_\text{(tj)}^{2}$ \citep{tjur_coefficients_2009}; these two metrics do not depend on the model's likelihood. Details about the functional forms of the additional measures can be found in the given references, also with a general overview found, for instance, in \cite{allison_what_2013}.
\begin{figure}[t!]
	\centering
	\includegraphics[scale=0.32]{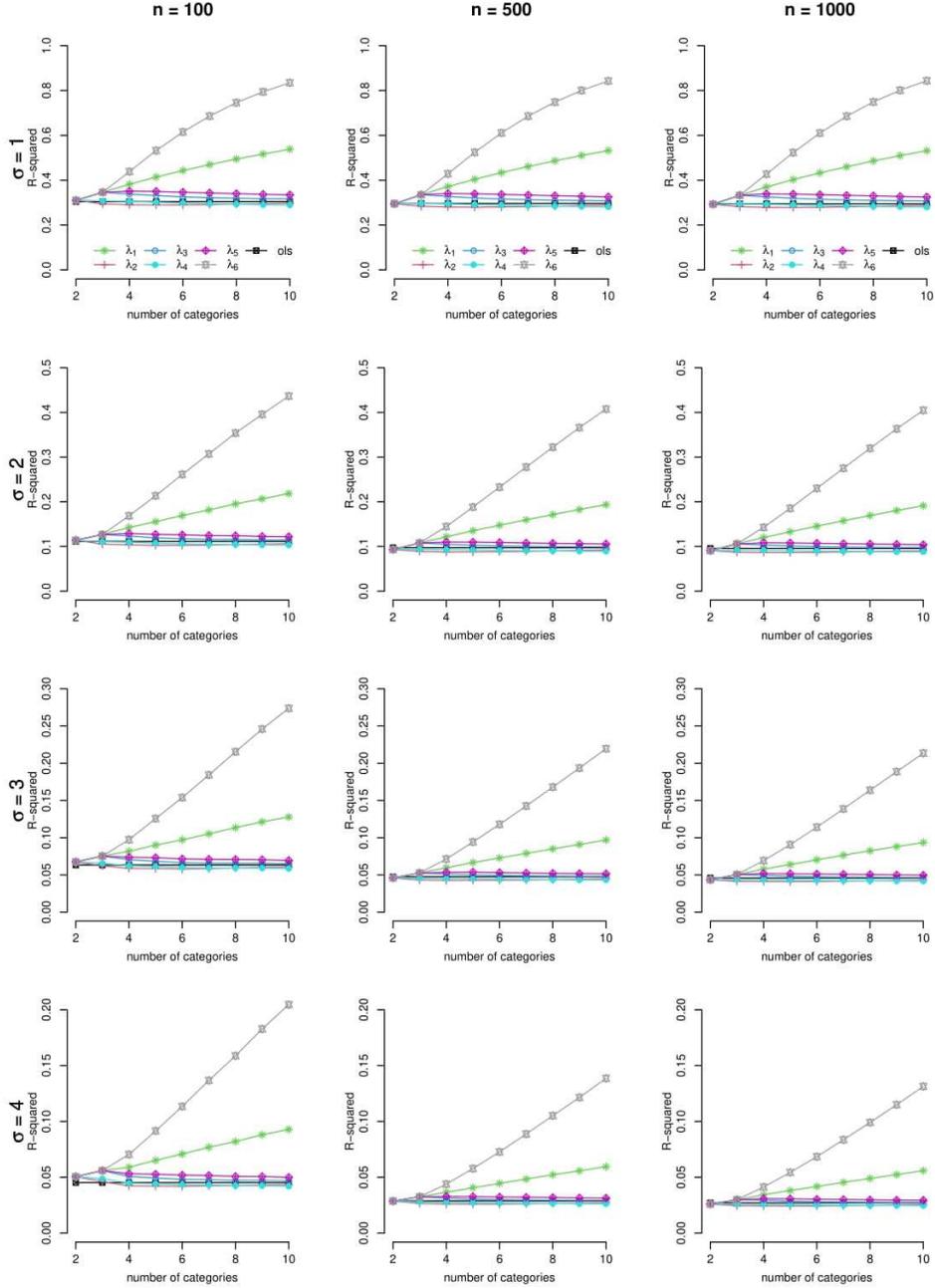}
	\vspace{3mm}
	\caption{\label{ordinal2} R-squared of multi-categorical response models against increasing number of response category ($r$ = $2, 3, \ldots, 10$), comparing the modified measure with the underlying measure (ols). Rows and columns respectively denote the four different specifications of standard deviation of $\tilde{y}_{i}$  and the three sample sizes.}
	\vspace{1.5ex}
\end{figure}

Given a thousand replications of all the measures under consideration, the corresponding errors $\Delta$ = $ R_{(r=2)}^{2}-R_\text{(ols)}^{2}$ were obtained. Figure \ref{error1} shows the extent to which these measures approximate the underlying $R_\text{(ols)}^{2}$. \!The modified $R_{(r = 2)}^{2}$ with $\lambda(2)$ = 2 (denoted by `ug2' in \mbox{Figures \ref{error1}} and \mbox{\ref{error2})} particularly performs very well, and also compares favorably with McKelvey \& Zavoina's $R_\text{(mz)}^{2}$, which reportedly approximated $R_\text{(ols)}^{2}$ very well in previous studies (see, e.g., \cite{hagle_goodness_1992}; \cite{windmeijer_goodness_1995}; \cite{veall_pseudo_1992}). The modified measure with $\tilde{\lambda}(2)$ = 3 (denoted by `ug3' in \mbox{Figures \ref{error1}} and \mbox{\ref{error2})}, however, over-estimates the underlying measure, whereas the rest of the measures under-estimate the same at different rates. In particular, when compared to McFadden's original $R_\text{(mf)}^{2}$, it is very obvious that the modified version (ug2) provides a reasonable improvement.

In order to  further examine the measures' performance, another five purely noise covariates (also drawn from i.i.d $U(0,1)$) were added to the models to fit, while the true data generating process as described above remains the same. Results are presented in Figure~\ref{error2}. However, there doesn't seem to be any substantial difference to the results presented in Figure~\ref{error1}.

\subsection{Results from ordinal models}\label{Sec:Ordinal}
We further investigate the performance of the modified measure in multi-categorical response models, also comparing it with the underlying and the existing measures. As earlier argued in Section~\ref{Sec:Latent}, in addition to approximating the underlying measure, a good summary measure for the discrete models should as well be invariant to the number of response categories $r$, used in the model, especially if the dependent variable $y_i$ is motivated by an underlying latent variable $\tilde{y}_{i}$. Thus, we obtained and compared the different summary measures under increasing $r=2,3,\ldots,10$. The non-likelihood measures were excluded at this point since they do not easily (or not at all) extend to multi-categorical models. Figure~\ref{ordinal1}, in particular, compares the existing measures with the underlying $R_\text{(ols)}^{2}$ (with each point being the average $R^2$ value over a 1000 replications), while Figure~\ref{ordinal2} compares the modification proposed to $R_\text{(ols)}^{2}$. In both plots, different sample sizes (columns) and variances (rows) are considered. As observed in Figure~\ref{ordinal1}, all the measures are somewhat affected by changes in the number of response categories, with $R_\text{(mf)}^{2}$ and $R_\text{(cs)}^{2}$ being affected the most. 
\noindent
Either of these two measures are widely reported in empirical studies for multi-categorical models, but as observed in Figure~\ref{ordinal1}, apart from not being invariant to $r$, they also diverge in opposite directions to each other as $r$ moves away from dichotomization, with $R_\text{(mf)}^{2}$, in particular, depreciating monotonically. In this instance, not only would the choice of $r$ mean two different conclusions for two independent researchers modeling the same predictor effects on the same dataset, choosing between any of these measures could as well lead to conflicting conclusions. On the contrary, the modified measure (Figures~\ref{ordinal2}) with certain penalty specifications provides a stable metric for determining the predictive strength of ordinal models. Even though effect-sizes in both Figures~\ref{ordinal1} and \ref{ordinal2} depreciate row-wise (i.e., as the latent error variance increases), the same shape is maintained comparing the different measures with the underlying measure. Overall, sample size doesn't seem to make any substantial difference. Of all the five penalties used for the modified measure, $\lambda_2$ and $\lambda_4$ seem to perform best. Both prove quite invariant to $r$ and also approximate the underlying measure closely.  Of course, the finding that $\lambda_2$ and $\lambda_4$ perform similarly is not surprising since both functions have quite similar shape on the $r$ values considered (see Figure~\ref{fig:penalty}). Overall, all non-linear candidate functions considered (i.e., $\lambda_2$--$\lambda_5$) appear to provide substantial improvement over the original McFadden $R_\text{(mf)}^{2}$.
%
\section{Empirical Application}\label{Sec:Empirical} 
For a practical application of the modified measure and comparison with related measures, an ordered multi-categorical response obtained via sensory evaluation of boar taint is considered. Due to animal welfare concerns, the production of entire male pigs is seen as a viable alternative to surgical castration. Elevated levels of so-called boar taint may, however, impair consumer acceptance (see, for example, \cite{trautmann_how_2014} and references therein). Boar taint is (presumably) caused by two malodorous volatile substances: Androstenone and skatole (compare, for example, \cite{MeierDinke_evaluating_2015})).

\begin{table}[t!]
	\footnotesize\centering\tabcolsep=7.5pt
	\caption{\label{sensory} Linear, binary and ordinal models of the sensory data, with androstenone (AN), skatole (SK) and interaction (AN:SK) as predictors of deviant smell}
	\vspace{0.5ex}
	\begin{tabular}{c|ccc|ccc|ccc}
		\hline\hline	
		& \multicolumn{3}{c|}{\textbf{Linear Model}}& \multicolumn{3}{c|}{\textbf{Binary Model}}& \multicolumn{3}{c}{\textbf{Ordinal Model}}\\
		\hline 
		&\textbf{B} & \textbf{SE-B} &  \textbf{Pr}($>|t|$)  &\textbf{B} & \textbf{SE-B} &  \textbf{Pr}($>|z|$) &\textbf{B} & \textbf{SE-B} &  \textbf{Pr}($>|z|$)\\
		\hline
		$\alpha$             & 1.853     &   0.023   &  0.00 ***  & $-$0.395  &  0.045   &  0.00 ***  &            &           &              \\ \hline
		$\alpha_1$           &           &           &            &           &          &            &  1.168     &   0.054   &  0.00 ***   \\ \hline
		$\alpha_2$           &           &           &            &           &          &            & $-$0.403   &   0.044   &  0.00 ***   \\ \hline
		$\alpha_3$           &           &           &            &           &          &            & $-$1.600   &   0.066   &  0.00 ***   \\ \hline
		$\alpha_4$           &           &           &            &           &          &            & $-$2.614   &   0.118   &  0.00 ***   \\ \hline
		AN                   & 0.179     &   0.024   &  0.00 ***  & 0.270     & 0.048    &  0.00 ***  &  0.240     &   0.038   &  0.00 ***   \\ \hline
		SK                   & 0.403     &   0.025   &  0.00 ***  & 0.482     & 0.051    &  0.00 ***  &  0.540     &   0.041   &  0.00 ***   \\ \hline
		AN:SK                & 0.092     &   0.020   &  0.00 ***  & 0.015     & 0.047    &  0.75      &  0.118     &   0.033   &  0.00 ***   \\
		\hline
		$R_\text{(cs)}^{2}$  &           &           &            &           &  0.188   &            &            &   0.302    &              \\ \hline
		$R_\text{(nk)}^{2}$  &           &           &            &           &  0.258   &            &            &   0.326    &              \\ \hline
		$R_\text{(mz)}^{2}$  &           &           &            &           &  0.293   &            &            &            &               \\ \hline
		$R_\text{(tj)}^{2}$  &           &           &            &           &  0.204   &            &            &            &              \\ \hline
		$R_\text{(mf)}^{2}$  &           &           &            &           &  0.159   &            &            &   0.139    &              \\ \hline
		$R_{(r)}^{2}: \lambda_1$  &           &           &            &           &  0.293   &            &            &   0.527    &             \\ \hline
		$R_{(r)}^{2}: \lambda_2$  &           &           &            &           &  0.293   &            &            &   0.377    &              \\ \hline
		$R_{(r)}^{2}: \lambda_3$  &           &           &            &           &  0.293   &            &            &   0.428    &              \\ \hline
		$R_{(r)}^{2}: \lambda_4$  &           &           &            &           &  0.293   &            &            &   0.392    &              \\ \hline
		$R_{(r)}^{2}: \lambda_5$  &           &           &            &           &  0.293   &            &            &   0.451    &             \\ \hline
		\mbox{$R_{(r)}^{2}: \lambda_6$}  &           &           &            &           &  0.293   &            &            &   0.807    &            \\ \hline
		$R_\text{(ols)}^{2}$      &           &  0.379    &            &           &          &            &            &            &             \\
		\hline\hline
		\multicolumn{10}{p{13.7cm}@{}}{Applicable R-squared for the different models are reported below each model. The significance code `***' indicates values $<$ 0.001.}
	\end{tabular}
	\vspace{-0.5ex}
\end{table}
In what follows, we consider data from an experimental study presented in \cite{morlein_interaction_2016}. (2016) where fat samples of more than a thousand samples of pig carcasses were collected and subjected to a thorough sensory evaluation and quantification using a panel of 10 trained assessors on a sensory score scale ranging from 0 = `untainted' to 5 = `strongly deviant smell'; also see M\"{o}rlein {\it et al}. (2021). The average panel rating for all the samples were also obtained and available for the present analysis.
The question of interest is how this measure is influenced by the samples' androstenone and skatole contents.  In practice, however, panel ratings are often discretized to a binary or multi-categorical variable, with a typical cut point for dichotomization (boar tainted/no boar taint) being 2; compare, for example, \cite{MeierDinke_evaluating_2015}. For multi-categorical grouping, \cite{morlein_interaction_2016} used the following subdivision of the average panel rating: $[0,1), [1,2), [2,3), [3,4)$ and $[4,5]$. We also adopt such grouping in our analysis of the olfactory perception of boar taint, fitting both a binary and ordinal probit model using androstenone, skatole and their interaction as predictors (compare \cite{morlein_interaction_2016} and \cite{Ugba_smoothing_2021}). Due to the skewed distribution of androstenone and skatole, the two covariates were standardized after being transformed logarithmically. In a few cases, however, androstenone had a value of zero, which may be due to androstenone content below the detection threshold, or defective measurement. Therefore, those observations were excluded from further analysis. A linear version of the categorical model having the same predictors, but with the average panel rating as response was fitted as well (similarly to \cite{morlein_interaction_2016}). As shown in Table~\ref{sensory}, comparable patterns of predictor effects are seen in all the models. The covariates, androstenone and skatole, are shown to be significant predictors of deviant smell $(p < 0.001)$ in all models. Interactions were also significant in all but the binary model. The overall goodness-of-fit of both the binary and ordinal model was assessed using the modified $R_{(r)}^{2}$ and other applicable measures, while the coefficient of determination $R_\text{(ols)}^{2}$ was calculated for the linear model. As shown in Table~\ref{sensory}, all the reported $R^{2}s$ indicate some association between the response and the predictors. As already observed in the simulation studies, the modified measure employing $\lambda_2$ or $\lambda_4$ appears to approximate $R_\text{(ols)}^{2}$ quite well, whereas particularly the convex penalty function above the diagonal ($\lambda_6$) does a very poor job if the number of levels increases. Although there is a drop in the numbers for all the measures in the binary case, the modified $R_{(r = 2)}^{2}$ and again McKelvey $\&$ Zavoina's $R_\text{(mz)}^{2}$ still perform best.

From the viewpoint of interpretation, particularly if the latent $R_\text{(ols)}^{2}$ is not available (which of course is typically the case in ordinal regression), the question remains whether the Pseudo $R^2$-values obtained amount to a substantial effect or not. First, we note that concerning McFadden's $R_\text{(mf)}^{2}$-values between 0.2 and 0.4 are taken to represent a very good fit of the model \citep{mcFadden_conditional_1974}. Simulations by \citep{domenich_urban_1975}. (1975) equivalence this range to 0.7 to 0.9 for a linear model \citep{louviere_stated_2000}. In a similar vein, when referencing an underlying measure, values of the modified measure within the latter range would represent very good fits. Thus, the overall goodness-of-fit of the ordinal model of the sensory data, considering $R_{(r)}^{2}$ (with either $\lambda_2$ or $\lambda_4$ penalty), indicate a moderately good fit. Further diagnostic checks via hypothesis tests could tell if lack of fit does exist or not. For instance, the tests suggested in \cite{Fagerland_test_2016}, see also \cite{jeong_goodness_2009}; \cite{yoo_statistical_2020}.

\section{Discussion}\label{Sec:Disc}
The R-squared measure is considered a very crucial diagnostic tool in empirical studies because it provides a quick evaluation of the predictive strength of the fitted models. However, whether or not to use a Pseudo R-squared measure for categorical response models has been an issue of intense debate in the literature for decades. A lot of measures have been proposed for this very purpose, with the very recent being measures proposed by \cite{zhang_coefficient_2017} for the generalized linear model and \cite{piepho_coefficient_2019} for the generalized linear mixed models. The latter, in particular, proposes a coefficient of determination that is defined on the linear predictor scale. Highlighting the pros and cons of several goodness-of-fit measures for the logistic regression model, \cite{allison_measures_2014} made mention of Tjur's coefficient of discrimination, denoted by $(R_\text{(tj)}^{2})$ in this article, for its simplicity and intuitive understanding, while also making a paradigm shift from recommending the Cox \& Snell $R_\text{(cs)}^{2}$ to the McFadden $R_\text{(mf)}^{2}$. A couple of reasons seem to support $R_\text{(mf)}^{2}$, particularly its simple formulation, base-rate stability in binary models, as well as, an intuitive interpretation as the proportional reduction in the log-likelihood statistics of fitted models \citep{menard_coefficients_2000}. Several statistical software products (SPSS and SAS, for instance) report $R_\text{(mf)}^{2}$ in their standard outputs for ordinal response models. Nevertheless, as observed in this study, we may not support such use of $R_\text{(mf)}^{2}$ in ordinal models. Apart from underestimating the underlying measure, and in a sharp contrast to similar measures which all appreciate towards the underlying measure under increasing number of response categories, $R_\text{(mf)}^{2}$ attaches smaller values to more complicated models (having a larger number of response categories) built on the same dataset. An alternative to $R_\text{(mf)}^{2}$ that redresses its key limitations is proposed in this study. In a nutshell, we recommend an exponentially penalized likelihood ratio index with a stabilizing penalty of $\lambda(r)$ = $\sqrt{2r}$, $\lambda(r)$ = $1 + \log_2{r}$, or a similar function. Results from simulation studies and real data examples very well attest to the usefulness of the proposed measure in binary and ordinal models. Our modification also provides a likelihood-based alternative to the McKelvey \& Zavoina $R_\text{(mz)}^{2}$, which is rarely reported due to its complexity in computation and interpretation. Finally, the modified measure is also supposed to mimic (at least to some extent) Rao's properties of the coefficient of determination mentioned earlier in this paper. Specifically, (1) it has a passably easy and intuitive interpretation as a penalized version of the proportional reduction in the $-$2 log-likelihood statistic, (2) it yields values that are between 0 and 1, (3) it is dimensionless, for instance, the scale of measurement of skatole and androstenone in the real data application is of no consequence to the modified $R^{2}$-values, and (4) as observed from the simulation studies, it also seems to be rather independent of the sample size.

The proposed measure and other goodness-of-fit measures for categorical models have been implemented in the R add-on package~\texttt{gofcat} \citep{ugba_serp_joss_2021}, available from the comprehensive R archive network (\href{https://CRAN.R-project.org/package=gofcat}{CRAN}). The sensory data \citep{morlein_interaction_2021} analyzed in this paper is available from \href{https://doi.org/10.5281/zenodo.4869352}{Zenodo}. Moreover, the generated comparison plots for simulation setting (b) in analogy to \mbox{Figures~\ref{error1}}--\ref{ordinal2} are available as part of an online appendix.

\section*{Acknowledgement}
This research was supported in part by Deutsche Forschungsgemeinschaft (DFG) through grant number GE2353/2-1.

\bibliographystyle{Chicago}
\bibliography{R2arxiv}

\begin{thebibliography}{}

\bibitem[\protect\citeauthoryear{Agresti}{Agresti}{1986}]{Agresti_applying_1986}
Agresti, A. (1986).
\newblock Applying ${R}^{2}-$type measures to ordered categorical data.
\newblock {\em Technometrics\/}~{\em 28}, 133--138.
\newblock doi: \url{10.2307/1270449}.

\bibitem[\protect\citeauthoryear{Agresti}{Agresti}{2002}]{agresti_categorical_2002}
Agresti, A. (2002).
\newblock {\em Categorical Data Analysis\/} (2nd ed.).
\newblock New York: John Wiley and Sons.
\newblock doi: \url{10.1002/0471249688}.

\bibitem[\protect\citeauthoryear{Allison}{Allison}{2013}]{allison_what_2013}
Allison, P. (2013).
\newblock What’s the best r-squared for logistic regression?
\newblock Available from: \url{https://statisticalhorizons.com/r2logistic}
  (accessed on 29-09-2021).

\bibitem[\protect\citeauthoryear{Allison}{Allison}{2014}]{allison_measures_2014}
Allison, P.~D. (2014).
\newblock Measures of fit for logistic regressions.
\newblock In {\em Proceedings of the SAS Global 2014 Conference}, Washington,
  DC, pp.\  1--12.

\bibitem[\protect\citeauthoryear{Cox and Snell}{Cox and
  Snell}{1989}]{cox_analysis_1989}
Cox, D.~R. and E.~J. Snell (1989).
\newblock {\em Analysis of Binary Data\/} (2nd ed.).
\newblock London: Chapman and Hall.

\bibitem[\protect\citeauthoryear{Domencich, McFadden, McFadden, and
  Associates}{Domencich et~al.}{1975}]{domenich_urban_1975}
Domencich, T., D.~McFadden, D.~McFadden, and C.~R. Associates (1975).
\newblock {\em Urban Travel Demand: A Behavioral Analysis : a Charles River
  Associates Research Study}.
\newblock Number v. 93 in Charles River Associates. Research Studies. Charles
  River Associates. Research Studies. Contributions to economic analysis, 93.
  North-Holland Publishing Company.
\newblock \url{https://books.google.de/books?id=VUZxnQEACAAJ}.

\bibitem[\protect\citeauthoryear{Fagerland and Hosmer}{Fagerland and
  Hosmer}{2016}]{Fagerland_test_2016}
Fagerland, M.~W. and D.~W. Hosmer (2016).
\newblock Tests for goodness of fit in ordinal logistic regression models.
\newblock {\em Journal of Statistical Computation and Simulation\/}~{\em 86},
  3398--3418.

\bibitem[\protect\citeauthoryear{Hagle and Mitchell~II}{Hagle and
  Mitchell~II}{1992}]{hagle_goodness_1992}
Hagle, T.~M. and G.~E. Mitchell~II (1992).
\newblock Goodness-of-fit measures for probit and logit.
\newblock {\em Am. J. Polit. Sci.\/}~{\em 36}, 762--784.
\newblock doi: \url{10.2307/2111590}.

\bibitem[\protect\citeauthoryear{Hauser}{Hauser}{1978}]{hauser_testing_1978}
Hauser, J.~R. (1978).
\newblock Testing the accuracy, usefulness, and significance of probabilistic
  choice models: An information-theoretic approach.
\newblock {\em Oper. Res.\/}~{\em 26\/}(3), 406--421.
\newblock doi: \url{10.1287/opre.26.3.406}.

\bibitem[\protect\citeauthoryear{Heinzl and {Mittlb{\"o}ck}}{Heinzl and
  {Mittlb{\"o}ck}}{2003}]{heinzl_pseudo_2003}
Heinzl, H. and M.~{Mittlb{\"o}ck} (2003).
\newblock Pseudo {R}-squared measures for poisson regression models with over-
  or underdispersion.
\newblock {\em Comput Stat Data Anal\/}~{\em 44}, 253--271.
\newblock doi: \url{10.1016/s0167-9473(03)00062-8}.

\bibitem[\protect\citeauthoryear{Hosmer and Lemeshow}{Hosmer and
  Lemeshow}{1989}]{hosmer_applied_1989}
Hosmer, D.~W. and S.~Lemeshow (1989).
\newblock {\em Applied Logistic Regression}.
\newblock New York: John Wiley and Sons.

\bibitem[\protect\citeauthoryear{Jeong and Lee}{Jeong and
  Lee}{2009}]{jeong_goodness_2009}
Jeong, K.~M. and H.~Y. Lee (2009).
\newblock Goodness-of-fit tests for the ordinal response models with
  misspecified links.
\newblock {\em Communications for Statistical Applications and Methods\/}~{\em
  16}, 697--705.

\bibitem[\protect\citeauthoryear{Long}{Long}{1997}]{long_regression_1997}
Long, J.~S. (1997).
\newblock {\em Regression Models for Categorical and Limited Dependent
  Variables}.
\newblock California: Sage Publications.

\bibitem[\protect\citeauthoryear{Louviere, Hensher, and Swait}{Louviere
  et~al.}{2000}]{louviere_stated_2000}
Louviere, J.~J., D.~A. Hensher, and J.~D. Swait (2000).
\newblock {\em Stated Choice Methods: Analysis and Application}.
\newblock Cambridge, UK: Cambridge University Press.

\bibitem[\protect\citeauthoryear{Maddala}{Maddala}{1983}]{maddala_limited_1983}
Maddala, G.~S. (1983).
\newblock {\em Limited-Dependent and Qualitative Variables in Econometrics}.
\newblock Cambridge University.

\bibitem[\protect\citeauthoryear{McFadden}{McFadden}{1974}]{mcFadden_conditional_1974}
McFadden, D. (1974).
\newblock Conditional logit analysis of qualitative choice behavior.
\newblock {\em Frontiers in Econometrics P. Zarembka (ed.)\/}, 105--142.

\bibitem[\protect\citeauthoryear{McKelvey and Zavoina}{McKelvey and
  Zavoina}{1976}]{mcKelvey_statistical_1976}
McKelvey, R.~D. and W.~Zavoina (1976).
\newblock A statistical model for the analysis of ordinal level dependent
  variables.
\newblock {\em J.Math. Sociol.\/}~{\em 4}, 103--120.

\bibitem[\protect\citeauthoryear{Meier-Dinkel, Gertheiss, M{\"{u}}ller, Wesoly,
  and {M{\"o}rlein}}{Meier-Dinkel et~al.}{2015}]{MeierDinke_evaluating_2015}
Meier-Dinkel, L., J.~Gertheiss, S.~M{\"{u}}ller, R.~Wesoly, and
  D.~{M{\"o}rlein} (2015).
\newblock Evaluating the performance of sensory quality control: the case of
  boar taint.
\newblock {\em Meat Sci.\/}~{\em 100}, 73--84.
\newblock doi: \url{10.1016/j.meatsci.2014.09.013}.

\bibitem[\protect\citeauthoryear{Menard}{Menard}{2000}]{menard_coefficients_2000}
Menard, S. (2000).
\newblock Coefficients of determination for multiple logistic regression
  analysis.
\newblock {\em Am. Stat.\/}~{\em 54}, 17--24.
\newblock doi: \url{10.1080/00031305.2000.10474502}.

\bibitem[\protect\citeauthoryear{{M{\"o}rlein}, Trautmann, Gertheiss,
  Meier-Dinkel, Fischer, Eynck, Heres, Looft, and Tholen}{{M{\"o}rlein}
  et~al.}{2016}]{morlein_interaction_2016}
{M{\"o}rlein}, D., J.~Trautmann, J.~Gertheiss, L.~Meier-Dinkel, J.~Fischer,
  H.-J. Eynck, L.~Heres, C.~Looft, and E.~Tholen (2016).
\newblock Interaction of skatole and androstenone in the olfactory perception
  of boar taint.
\newblock {\em Journal of Agricultural and Food Chemistry\/}~{\em 64},
  4556--4565.
\newblock doi: \url{10.1021/acs.jafc.6b00355}.

\bibitem[\protect\citeauthoryear{{M{\"o}rlein}, Trautmann, Gertheiss,
  Meier-Dinkel, Fischer, Eynck, Heres, Looft, and Tholen}{{M{\"o}rlein}
  et~al.}{2021}]{morlein_interaction_2021}
{M{\"o}rlein}, D., J.~Trautmann, J.~Gertheiss, L.~Meier-Dinkel, J.~Fischer,
  H.-J. Eynck, L.~Heres, C.~Looft, and E.~Tholen (2021).
\newblock Androstenone, skatole and the olfactory perception of boar taint
  (1.0.0) [data set].
\newblock doi: \url{10.5281/zenodo.4869352}.

\bibitem[\protect\citeauthoryear{Nagelkerke}{Nagelkerke}{1991}]{nagelkerke_note_1991}
Nagelkerke, N. J.~D. (1991).
\newblock A note on a general definition of the coefficient of determination.
\newblock {\em Biometrika\/}~{\em 78}, 691--692.
\newblock doi: \url{10.1093/biomet/78.3.691}.

\bibitem[\protect\citeauthoryear{Piepho}{Piepho}{2019}]{piepho_coefficient_2019}
Piepho, H.~P. (2019).
\newblock A coefficient of determination (${R}^{2}$) for generalized linear
  mixed models.
\newblock {\em Biometrical Journal\/}, 1--13.
\newblock doi: \url{10.1002/bimj.201800270}.

\bibitem[\protect\citeauthoryear{Rao}{Rao}{1973}]{rao_linear_1973}
Rao, C.~R. (1973).
\newblock {\em Linear Statistical Inference and its Applications\/} (2nd ed.).
\newblock New York: Wiley.

\bibitem[\protect\citeauthoryear{{RC Team}}{{RC Team}}{2022}]{R_language_2022}
{RC Team} (2022).
\newblock R: A language and environment for statistical computing.
\newblock Vienna, Austria. \url{https://www.R-project.org/}.

\bibitem[\protect\citeauthoryear{Tjur}{Tjur}{2009}]{tjur_coefficients_2009}
Tjur, T. (2009).
\newblock Coefficients of determination in logistic regression models-a new
  proposal: the coefficient of discrimination.
\newblock {\em Am. Stat.\/}~{\em 63}, 366--372.
\newblock doi: \url{10.1198/tast.2009.08210}.

\bibitem[\protect\citeauthoryear{Trautmann, Gertheiss, Wicke, and
  {M{\"o}rlein}}{Trautmann et~al.}{2014}]{trautmann_how_2014}
Trautmann, J., J.~Gertheiss, M.~Wicke, and D.~{M{\"o}rlein} (2014).
\newblock How olfactory acuity affects the sensory assessment of boar fat: a
  proposal for quantification.
\newblock {\em Meat Sci.\/}~{\em 98}, 255--262.
\newblock doi: \url{10.1016/j.meatsci.2014.05.037}.

\bibitem[\protect\citeauthoryear{Ugba}{Ugba}{2021}]{ugba_serp_joss_2021}
Ugba, E.~R. (2021).
\newblock serp: An {R} package for smoothing in ordinal regression.
\newblock {\em Journal of Open Source Software\/}~{\em 6\/}(66), 3705.
\newblock doi: \url{10.21105/joss.03705}.

\bibitem[\protect\citeauthoryear{Ugba and Gertheiss}{Ugba and
  Gertheiss}{2018}]{ugba_augmented_2018}
Ugba, E.~R. and J.~Gertheiss (2018).
\newblock An augmented likelihood ratio index for categorical response models.
\newblock Bristol, UK, pp.\  293--298.

\bibitem[\protect\citeauthoryear{Ugba, {M{\"o}rlein}, and Gertheiss}{Ugba
  et~al.}{2021}]{Ugba_smoothing_2021}
Ugba, E.~R., D.~{M{\"o}rlein}, and J.~Gertheiss (2021).
\newblock Smoothing in ordinal regression: An application to sensory data.
\newblock {\em Stats\/}~{\em 4}, 616--633.
\newblock doi: \url{10.3390/stats4030037}.

\bibitem[\protect\citeauthoryear{Veall and Zimmermann}{Veall and
  Zimmermann}{1992}]{veall_pseudo_1992}
Veall, M.~R. and K.~F. Zimmermann (1992).
\newblock Pseudo-${R}^2$'s in the ordinal probit model.
\newblock {\em J. Math. Sociol.\/}~{\em 4}, 103--120.
\newblock doi: \url{10.1080/0022250x.1992.9990094}.

\bibitem[\protect\citeauthoryear{Veall and Zimmermann}{Veall and
  Zimmermann}{1996}]{Veal_pseudo_1996}
Veall, M.~R. and K.~F. Zimmermann (1996).
\newblock Pseudo-${R}^{2}$ measures for some common limited dependent variable
  models.
\newblock {\em Journal of Economic Surveys\/}~{\em 10}, 241--259.
\newblock doi: \url{10.1111/j.1467-6419.1996.tb00013.x}.

\bibitem[\protect\citeauthoryear{Windmeijer}{Windmeijer}{1995}]{windmeijer_goodness_1995}
Windmeijer, F. A.~G. (1995).
\newblock Goodness-of-fit measures in binary choice models.
\newblock {\em Econom. Rev.\/}~{\em 14}, 101--116.
\newblock doi: \url{10.1080/07474939508800306}.

\bibitem[\protect\citeauthoryear{Yoo and Kim}{Yoo and
  Kim}{2020}]{yoo_statistical_2020}
Yoo, M. and D.~Kim (2020).
\newblock Statistical tests for biosimilarity based on relative distance
  between follow-on biologics for ordinal endpoints.
\newblock {\em Communications for Statistical Applications and Methods\/}~{\em
  22}, 1--14.
\newblock doi: \url{10.29220/CSAM.2020.27.1.001}.

\bibitem[\protect\citeauthoryear{Zhang}{Zhang}{2017}]{zhang_coefficient_2017}
Zhang, D. (2017).
\newblock A coefficient of determination for generalized linear models.
\newblock {\em The American Statisticians\/}~{\em 71}, 310--316.
\newblock doi: \url{10.1080/00031305.2016.1256839}.

\end{thebibliography}
\end{document}